\documentstyle[12pt]{article}
\def\nab#1{{\nabla_{#1}}}
\def\nabstar#1{\nabla\kern-0.5pt\smash{\raise 4.5pt\hbox{$\ast$}}
               \kern-4.5pt_{#1}}

\def\drvstar#1{\partial\kern-0.5pt\smash{\raise 4.5pt\hbox{$\ast$}}
               \kern-5.0pt_{#1}}

\def\newline{\relax\ifhmode\null\hfil\break\else\nonhmodeerr@\newline\fi}
\def\frac#1#2{{#1\over#2}}
\def\text#1{{\hbox{\rm #1}}}
\def\flushpar{{\par \noindent}}

\newcommand{\beq}{\begin{equation}}
\newcommand{\eeq}{\end{equation}}
\newcommand{\bea}{\begin{eqnarray}}
\newcommand{\eea}{\end{eqnarray}}
\def\Id{ \mbox{1\hspace{-1.2mm}I} }
\def\BE{\begin{equation}}
\def\EE{\end{equation}}
\def\BA{\begin{eqnarray}}
\def\EA{\end{eqnarray}}
\def\BAN{\begin{eqnarray*}}
\def\EAN{\end{eqnarray*}}

\def\nn{\nonumber\\}

\def\tr{\mbox{tr}}

\def\gm5{\gamma^5}

\def\bpsi{\bar{\psi}}

\begin{document}
\thispagestyle{empty}
\begin{flushright}
NTUTH-98-090 \\
September 1998
\end{flushright}
\bigskip\bigskip\bigskip
\vskip 2.5truecm
\begin{center}
{\LARGE {The axial anomaly of Ginsparg-Wilson fermion}}
\end{center}
\vskip 1.0truecm
\centerline{Ting-Wai Chiu}
\vskip5mm
\centerline{Department of Physics, National Taiwan University}
\centerline{Taipei, Taiwan 106, R.O.C}
\centerline{\it E-mail : twchiu@phys.ntu.edu.tw}
\vskip 2cm
\bigskip \nopagebreak \begin{abstract}
\noindent

The axial anomaly of Ginsparg-Wilson fermion operator $ D $ is discussed
in general for the operator $ R $ which enters the chiral
symmetry breaking part in the Ginsparg-Wilson relation. The axial anomaly
and the index of $ D $ as well as the exact realization of the Atiyah-Singer
index theorem on the lattice are determined solely by the topological
characteristics of the chirally symmetric operator $ D_c $ in the chiral
limit $ R \to 0 $.


\end{abstract}
\vskip 1.5cm

\newpage\setcounter{page}1

Ginsparg-Wilson (GW) relation \cite{gwr} has been playing an important role
in recent theoretical developments of the chiral symmetry on the lattice
\cite{hn98:8,nieder98:7}.
It was derived by Ginsparg and Wilson in 1981 as
the remnant of chiral symmetry on the lattice after blocking a chirally
symmetric theory with a chirality breaking local renormalization group
transformation. The GW relation for exactly massless Dirac fermion
operator $ D $ is
\beq
\label{eq:gwo}
D \gamma_5 + \gamma_5 D = 2 D \gamma_5 R D,
\eeq
where the chiral symmetry breaking on the RHS of Eq. (\ref{eq:gwo})
constitutes an invertible hermitian operator $ R $
which is local in the position space and trivial in the Dirac space.

In topologically trivial background gauge field, $ D^{-1} $ exists
and Eq. (\ref{eq:gwo}) can be written in terms of the fermion propagator
$ D^{-1} $
\beq
D^{-1} \gamma_5 + \gamma_5 D^{-1} = 2 \gamma_5 R
\label{eq:gwi}
\eeq
This equation displays explicitly that the chirality breaking part in the
GW fermion propagator is dictated by the local operator $ R $,
in contrast to those non-local breakings due to adding a chiral symmetry
breaking term explicitly to the action.
 
The GW relation, Eq. (\ref{eq:gwo}) can be rewritten as
\beq
\label{eq:gwc}
D \gamma_5 ( \Id - R D ) + ( \Id - D R ) \gamma_5 D = 0
\eeq
Then it becomes evident that the lattice action
$ A = \bar\psi D \psi $ is invariant under the exact chiral transformation
on the lattice \cite{ml98:2} :
\bea
\label{eq:lus1}
\psi &\rightarrow& \exp [ i \theta  \gamma_5 ( \Id - R D ) ] \psi, \\
\label{eq:lus2}
\bar\psi &\rightarrow& \bar\psi \exp [ i \theta (\Id- D R) \gamma_5 ],
\eea
where $ \theta $ is a global parameter.
Consequently the axial anomaly can be deduced from the change of fermion
integration measure under the exact chiral transformation
(\ref{eq:lus1})-(\ref{eq:lus2}), and its sum over all sites is equal
to the index of $ D $ which is a well defined integer \cite{ph98:1,ml98:2}.
The Ginsparg-Wilson fermion circumvents the Nielson-Ninomiya no-go theorem
\cite{no-go} by breaking the continuum chiral symmetry
$ \{ \gamma_5, D \}=0 $
but keeping the exact chiral symmetry on the lattice, Eq. (\ref{eq:gwc}),
hence $ D $ can be constructed to be local and free of species doubling.

The salient feature of the GW relation Eq. (\ref{eq:gwo}) in
topologically nontrivial background gauge field is that its chiral
symmetry breaking part does not have any effects on the zero modes.
This ensures that $ D $ has exact zero modes with definite
chiralities. Consequently the axial anomaly and the index of $ D $ are
invariant with respect to $ R $, and thus they
are determined solely by the chirally symmetric operator $ D_c $ in
the chiral limit $ R \to 0 $, where the GW relation is completely turned
off. The role of the GW relation is to transform the nonlocal and sometimes
singular ( i.e., divergent ) $ D_c $ into a class of local and well defined
fermion operators $D$ while keeping the topological characteristics of $ D_c $
invariant. If we have a $ D_c $ which can possess nonzero indices, then any
$ D $ constructed from this $ D_c $ using the general solution of
GW relation has the same topological characteristics.
Conversely, if we have a Ginsparg-Wilson $ D $ which has proper indices in
topologically nontrivial gauge fields, then there must exist a chirally
symmetric $ D_c $ which has the same topological characteristics of $ D $,
and we can use this $ D_c $ to construct a class of GW fermion operators.
The purpose of this paper is to discuss the $R$-invariance of the
axial anomaly, the index and the Atiyah-Singer index theorem on the
lattice, and show that their realizations on the lattice are actually
beyond the control of the GW relation but determined solely by the
topological characteristics of the chirally symmetric $ D_c $ in the
chiral limit $ R \to 0 $. For the axial anomaly recently constructed by
L\"uscher \cite{ml98:8}, we show that the term proportional to
$ F \tilde{F} $ is independent of $ R $ but depends on the topological
characteristics of $ D_c $, while the divergence term may depend on $ R $
in a very complicated way, but we show that its relationship to the
divergence of axial vector current is $R$-invariant.

We begin by reviewing the general solution of the Ginsparg-Wilson
relation which has been discussed in ref. \cite{twc98:6a}.
Our purpose here is to set up the notations.
The general solution of Eq. (\ref{eq:gwo}) can be
formally represented by
\beq
D = D_c ( \Id + R D_c )^{-1} = ( \Id + D_c R )^{-1} D_c
\label{eq:gwf}
\eeq
where $ D_c $ is the chirally symmetric fermion operator of
Eq. (\ref{eq:gwo}) in the chiral limit $ R \to 0 $,
\beq
D_c \gamma_5 + \gamma_5 D_c = 0.
\label{eq:Dc}
\eeq
It is instructive to show that these two different forms of the general
solution in (\ref{eq:gwf}) are equal and they satisfy Eq. (\ref{eq:gwo}).
The details are shown in the following.
\BAN
& & D_c ( \Id + R D_c )^{-1} \\
&=& ( \Id + D_c R )^{-1} ( \Id + D_c R ) D_c ( \Id + R D_c )^{-1} \\
&=& ( \Id + D_c R )^{-1} D_c ( \Id + R D_c ) ( \Id + R D_c )^{-1} \\
&=& ( \Id + D_c R )^{-1} D_c
\EAN
Using the general solution (\ref{eq:gwf}) to substitute $ D $ on the RHS of
Eq. (\ref{eq:gwo}), and inserting Eq. (\ref{eq:Dc}), we obtain
\BAN
& & 2 D \gamma_5 R D       \\
&=& 2 (\Id + D_c R )^{-1} D_c \gamma_5 R D_c (\Id + R D_c )^{-1}  \\
&=& (\Id + D_c R )^{-1} D_c R \gamma_5 D_c (\Id + R D_c )^{-1} +  \\
& & (\Id + D_c R )^{-1} D_c \gamma_5 R D_c (\Id + R D_c )^{-1} +  \\
& & (\Id + D_c R )^{-1} ( \gamma_5 D_c + D_c \gamma_5 )(\Id + R D_c )^{-1} \\
&=& (\Id + D_c R )^{-1} ( \Id + D_c R ) \gamma_5 D_c (\Id + R D_c )^{-1} + \\
& & (\Id + D_c R )^{-1} D_c \gamma_5 ( \Id + R D_c ) (\Id + R D_c )^{-1} \\
&=& \gamma_5 D + D \gamma_5
\EAN
For any $ D_c $ ( hence $ D $ ) satisfying the hermiticity condition
\bea
\gamma_5 D_c \gamma_5 = D_c^{\dagger},
\label{eq:hermit}
\eea
then $ D_c $ is antihermitian and it is evident from \cite{twc98:6a} that
there is one to one correspondence \footnote{This observation is
made by Herbert Neuberger in a private communication to the author.}
between $ D_c $ and a unitary operator $ V $ satisfying the hermiticity
condition ( $ \gamma_5 V \gamma_5 = V^{\dagger} $ )
\bea
D_c = (\Id + V )(\Id - V )^{-1}, \hspace{4mm} V = (D_c - \Id)( D_c + \Id)^{-1}.
\label{eq:VDc}
\eea
It is interesting to note that $ V $ exists for all $ D_c $, in particular,
when $ D_c \to \pm i \infty $, $ V \to +1 $;
but when $ D_c \to 0 $, $ V \to -1 $. The eigenvalues of $ V $ fall on
a unit circle with center at zero. On the other hand, $ D_c $ becomes
singular ( with simple poles ) when $ V $ has real eigenvalues $ +1 $. 
This is exactly what we want it to happen when the background gauge field is
topologically nontrivial. However $ D $ is still well defined even when
$ D_c $ is singular. Substituting $ D_c $ into (\ref{eq:gwf}),
we obtain the general solution of $ D $ in terms of a unitary operator
$ V $ satisfying the hermiticity condition \cite{twc98:6a},
\bea
\label{eq:gwsola}
D = ( \Id + V )[ ( \Id - V ) + R ( \Id + V ) ]^{-1} \\
\label{eq:gwsolb}
  = [ (\Id - V ) + ( \Id + V ) R ]^{-1} ( \Id + V ).
\eea
The general solution Eqs. (\ref{eq:gwsola}), (\ref{eq:gwsolb}) can be
casted into many different forms, for example \cite{nieder98:7},
\bea
D = \frac{1}{\sqrt{2R}} ( \Id + W ) \frac{1}{\sqrt{2R}}
\label{eq:nieder}
\eea
where $ W $ is a unitary operator which can be expressed in terms of
the $ V $ and $ R $. However, since $ W $ depends on $ R $, using
this form could easily obscure some simple facts especially in the chiral
limit $ R \to 0 $. Therefore we refrain from using Eq. (\ref{eq:nieder})
for analytic studies involving the general solution of the GW relation.

In general, the chirally symmetric operator $ D_c $ is constructed such
that it agrees with the continuum Dirac operator
$ \gamma_\mu ( \partial_\mu + i A_\mu ) $ in the classical continuum limit.
It is obvious from Eq. (\ref{eq:gwf}) that if $ D_c $ has species doubling,
$ D $ must have species doubling too. In order to avoid species doubling
for a chirally symmetric operator, $ D_c $ must be nonlocal.
We refer to ref. \cite{twc98:6b} for a few explicit examples of $ D_c $.
Next we consider the topological characteristics of a Dirac operator $ D $
by comparing its index, $ \mbox{index}(D) $ to the topological charge $ Q $
of the smooth background gauge field. The index of $ D $ is defined as
$ N_{-} - N_{+} $ where $ N_{+} ( N_{-} ) $ denotes the number of zero modes
of $ D $ with positive ( negative ) chirality. If $ \mbox{index}(D) = 0 $
for any smooth gauge configurations, then $ D $ is called
{\it topologically trivial.}
If $ \mbox{index}(D) = Q $, then $ D $ is called {\it topologically proper.}
If $ D $ is not topologically trivial but $ \mbox{index}(D) \ne Q $,
then $ D $ is called {\it topologically improper.} Examples of these
three different cases are given below and detailed discussions in
ref. \cite{twc98:9c}. From Eq. (\ref{eq:gwf}), it is obvious that
any zeromode of $ D $ is a zeromode of $ D_c $ and vice versa.
{\it Therefore the index of $ D $ is the same as the index of $ D_c $,
     independent of $ R $. }
\bea
\mbox{index} [D,A] = \mbox{index} [D_c,A] = N_{-}[D_c,A] - N_{+}[D_c,A]
\label{eq:index}
\eea
where we have indicated explicitly that the number of zero modes and
the index are functionals of $ D_c $ ( $ D $ ) and the gauge field $ A $.
However, {\it if $ D_c $ is well defined, its index must be zero. }
This can be easily proved in the following. \newline
{\it Proof : }   Consider $ R = r \Id $, then
$ \mbox{index} (D) = r \sum_x tr[ \gamma_5 D(x,x) ] $ \cite{ph98:1,ml98:2}.
In the limit $ r \to 0 $, $ D \to D_c $,
if $ D_c $ is well defined, then $ \sum_x tr[ \gamma_5 D(x,x) ] $ is finite,
thus multiplying $ r $ gives $ \mbox{index} (D) = 0 $.
This completes the proof. \\
{\it Therefore in order for $ D_c $ to possess non-zero indices,
$ D_c $ must be singular
( or equivalently, $ V $ has real eigenvalue
pairs $ \pm 1 $ ) in topologically nontrivial background gauge field }
\cite{twc98:6a}. \\
To summarize above discussions, we list
the necessary requirements for $ D_c $ to enter Eq. (\ref{eq:gwf})
for constructing a local and well defined $ D $ such that $ D $ could
possibly reproduce the continuum physics :
\begin{description}
\item[(a)] $ D_c $ is chirally symmetric and agrees with
    $ \gamma_\mu ( \partial_\mu + i A_\mu ) $ in the
    classical continuum limit.
\item[(b)] $ D_c $ is free of species doubling.
\item[(c)] $ D_c $ is nonlocal.
\item[(d)] $ D_c $ is well defined in topologically trivial background
             gauge field.
\item[(e)] $ D_c $ is singular ( or equivalently, $ V $ has real
             eigenvalue pairs $ \pm 1 $ ) in topologically non-trivial
             background gauge field.
\end{description}
It should be noted that each one of these constraints is required
to be satisfied. For example, if (b) is not satisfied, then the index of
$ D_c $ must be incorrect no matter other constraints are satisfied or not.
The most difficult task is to find a $ D_c $ satisfying (e).

At this point, it is instructive to consider examples of topologically
proper, trivial and improper $ D_c $ respectively.
First we consider the Neuberger-Dirac operator
\cite{hn97:7} in the chiral limit ( $ R \to 0 $ )
\bea
D_c = 2 M \frac{ \Id + V }{ \Id - V }, \hspace{4mm}
V = D_w ( D_w^{\dagger} D_w )^{-1/2}
\label{eq:DcV}
\eea
where $ M $ is an arbitrary mass scale, $ D_w $ is the Wilson-Dirac fermion
operator with negative mass $ -m_0 $ and Wilson parameter $ r_w > 0 $
\bea
D_w = - m_0
      + \frac{1}{2} [ \gamma_{\mu} ( \nabstar{\mu} + \nab{\mu} ) -
                      r_w  \nabstar{\mu} \nab{\mu} ]
\eea
where $ \nab{\mu} $ and $ \nabstar{\mu} $ are the forward and
backward difference operators defined in the following,
\bea
\nab{\mu}\psi(x) &=& \
   U_\mu(x)\psi(x+\hat{\mu})-\psi(x)  \nonumber \\
\nabstar{\mu} \psi(x) &=& \ \psi(x) -
   U_\mu^{\dagger}(x-\hat{\mu}) \psi(x-\hat{\mu})  \nonumber
\eea
Using Eq. (\ref{eq:gwf}), we obtain the generalized Neuberger-Dirac operator
\bea
D = 2 M ( \Id + V ) [ ( \Id - V ) + 2 M R ( \Id + V ) ]^{-1}
\eea
For $ m_0 \in ( 0, 2 r_w ) $, both $ D $ and $ D_c $ are topologically
proper and reproduce the exact index theorem on a finite lattice for
smooth background gauge fields as first demonstrated in
\cite{twc98:4,twc98:9b}. On the other hand, for $  m_0 \le 0 $,
that is, $ D_w $ with a positive mass, then $ D $ and $ D_c $ becomes
topologically trivial \cite{hn98:1}. We note that in both trivial and
proper cases $ D $ is free of species doubling and is local for properly
chosen $ R $. For $ m_0 \ge 2 r_w $,
both $ D $ and $ D_c $ are either topologically improper or trivial.
We emphasize that these ranges of $ m_0 $ values depend on the gauge
configurations, and the ranges given above are exact only in the free
field limit. We refer to ref. \cite{twc98:9c} for further discussions
on the topological characteristics of the Neuberger-Dirac operator.

It is interesting to note that {\it the GW relation does not play any role
in determining the index of $ D_c $.} The GW relation only ensures that the
chiral symmetry is broken in such a delicate way that the exact zero modes
of $ D_c $ are also exact zero modes of $ D $ with the same chiralities,
and thus the index of $ D $ is equal to the index of $ D_c $.
The role of $ R $ is to convert the nonlocal and sometimes singular $ D_c $
into a local and well defined $ D $, as well as to determine the locality
of $ D $. If one could not find any $ D_c $ satisfying all constraints
(a)-(e), then the GW relation does not have much use at all.
On the other hand, if one knows one $ D_c $ satisfying all constraints (a)-(e),
then the resulting $ D $ must have exact chiral symmetry on the lattice as
well as all attractive features pointed out in \cite{ph98:2,chand98:5}.

The fermion integration measure in general is not invariant under the
exact chiral transformation (\ref{eq:lus1})-(\ref{eq:lus2}), and
this gives the axial anomaly on the lattice \cite{ml98:2}
\bea
q(x,A;D) = \tr[ \gamma_5 ( R D ) (x,x) ]
\label{eq:qxR}
\eea
where $ \tr $ denotes the trace over Dirac indices.
We note that $ q(x,A;D) $ is a function of $ x $ and the gauge field
$ A_\mu $ in the neighborhood around $ x $, and a functional of $ D $.
The sum of $ q(x,A;D) $ over all lattice sites is shown
\cite{ph98:1,ml98:2,hn97:7} to equal to the index of $ D $
\bea
\sum_x q(x,A;D) &=& \sum_x \tr[ \gamma_5 ( R D )(x,x) ] \nn
&=& N_{-}[D,A] - N_{+}[D,A] \equiv \mbox{index} [D,A]
\label{eq:sum_q}
\eea
where $ N_{+} ( N_{-} ) $ denotes the number of zero modes of
$ D $ with positive ( negative ) chirality.
In this paper, we restrict our discussions
to $ U(1) $ gauge theory with single fermion flavor.
Since the index of $ D $ is independent of $ R $ [ Eq. (\ref{eq:index}) ],
the total axial anomaly must be invariant under any smooth deformations
of $ R $
\bea
\delta_R \sum_x q(x,A;D) = 0
\label{eq:del_qxR}
\eea
Recently L\"uscher \cite{ml98:8} has proved that for $ R = 1/2 $,
if the axial anomaly $ q(x,A;D) $ satisfies
\bea
\sum_x \delta q(x,A;D) = 0
\label{eq:local}
\eea
for any local deformations of the gauge field, then
\bea
q(x,A;D) = \gamma \ c[D] \ \epsilon_{\mu\nu\lambda\sigma} F_{\mu\nu} (x)
                                    F_{\lambda\sigma}(x+\hat\mu+\hat\nu)
                 + \partial_\mu G_\mu(x,A;D)
\label{eq:qFkR}
\eea
where $ \gamma $ is a constant, $ c[D] $ is a functional of $ D $ which
only depends on the topological characteristics of $ D $, and
\bea
\partial_\mu G_\mu(x,A;D) =
\sum_\mu \left[ G_\mu(x,A;D) - G_\mu(x-\hat\mu,A;D) \right].
\eea
The explicit form of the current $ G_\mu(x,A;D) $ is supposed to be
very complicated.
We note that $ c[D] $ is not shown explicitly in \cite{ml98:8}.
Since the total axial anomaly is invariant under smooth deformations of
$ R $, Eq. (\ref{eq:del_qxR}), as well as for any local deformations of the
gauge field, Eq. (\ref{eq:local}), then the theorem proved by
L\"uscher \cite{ml98:8} can be applied to the case of the general $ R $
\footnote{This is confirmed by Martin L\"uscher in a
private communication to the author.},
and the axial anomaly can be written in the same form as Eq. (\ref{eq:qFkR})
on an infinite lattice. But on a finite lattice, $ q(x,A;D) $ must be modified
by the boundary effects. However, if $ D $ is local, the boundary
effects are negligible, then Eq. (\ref{eq:qFkR}) can be used for the axial
anomaly on a finite lattice. Since the locality of $ D $ is
controlled by $ R $, we can infer that all boundary effects must have
dependence on $ R $ and hence they do not enter the $ F \tilde{F} $ term.
[ As we will show later, the $ F \tilde{F} $ term does not depend on $ R $. ]
Furthermore they must be in the form of the divergence term,
otherwise Eq. (\ref{eq:qFkR}) cannot be consistent with Eq. (\ref{eq:divJ5}),
the divergence of axial vector current on a finite lattice.
Therefore we conclude that Eq. (\ref{eq:qFkR}) is applicable for
a finite lattice as well as for an infinite lattice, and the boundary
effects only enter the divergence term.
Now we proceed our discussions for a finite lattice embedded in a
4-dimensional torus. After summing $ q(x,A;D) $ over all sites,
the divergence term is removed, then Eq. (\ref{eq:sum_q}) gives
\bea
\gamma \ c[D] \ \sum_x \epsilon_{\mu\nu\lambda\sigma} F_{\mu\nu} (x)
                            F_{\lambda\sigma}(x+\hat\mu+\hat\nu)    \nn
= N_{-}[D,A] - N_{+}[D,A] = \mbox{index} [D,A]
\label{eq:index_thm}
\eea
Since $ \mbox{index} [D,A] $ is independent of $ R $ [ Eq. (\ref{eq:index}) ],
then Eq. (\ref{eq:index_thm}) implies that $ c[D] $ must be also
independent of $ R $,
\bea
c[D]=c[D_c]
\label{eq:gDc}
\eea
So we have proved that all $ R $ dependence in $ q(x,A;D) $ only
resides in the current $ G_\mu(x,A;D) $.
Then Eq. (\ref{eq:index_thm}) can be rewritten in the following.
\bea
\gamma \ c[D_c] \ \sum_x \epsilon_{\mu\nu\lambda\sigma} F_{\mu\nu} (x)
                            F_{\lambda\sigma}(x+\hat\mu+\hat\nu)  \nn
= N_{-}[D_c,A] - N_{+}[D_c,A] = \mbox{index} [D_c,A]
\label{eq:index_thm_a}
\eea
Since $ \mbox{index} [D_c,A] = 0 $ if $ D_c $ is topologically trivial, then
$ c[D_c] $ also vanishes for trivial $ D_c $.
Now we consider two topologically proper $ D_c $ which are identical
except their electric charges of the fermion field.
Then for a given topologically non-trivial gauge configuration,
Eq. (\ref{eq:index_thm_a}) implies that the ratio of $ c[D_c]'s $
corresponding to these two $ D_c $ must equal to the ratio of
two integers corresponding to their indices.
It follows from this that $ c[D_c] $ must be an integer.
Therefore we can fix $ c[D_c] = 1 $ for electric charge equal to one and
then determine the value of the constant $ \gamma $ by evaluating
$ \sum_x \epsilon_{\mu\nu\lambda\sigma} F_{\mu\nu} (x)
                            F_{\lambda\sigma}(x+\hat\mu+\hat\nu) $
on the LHS for the simplest nontrivial gauge configuration with constant
field tensor, namely, $ F_{12} = 2 \pi /(L_1 L_2 )$ and
$ F_{34} = 2 \pi / ( L_3 L_4 ) $
while other $ F's $ are zero. After restoring the unit of electric charge,
we obtain $ \gamma $ to be integer multiple of $ e^2/(32 \pi^2) $ for
$ U(1) $ gauge theory with single fermion flavor.
For Neuberger-Dirac fermion operator \cite{hn97:7}, this integer is
determined to be unity by a recent perturbation calculation of
$ q(x,A;D) $ \cite{kiku98:6}.
For the general Ginsparg-Wilson fermion with $ R = r \Id $,
and $ D_c $ satisfying constraints (a)-(e) \footnote{ In ref. \cite{twc98:10a},
it is only required that the chiral fermion operator $ D_c(p) $
is free of species doubling and in the
free field limit behaves like $ i \gamma_\mu p_\mu $ as $ p \to 0 $. },
our recent perturbation calculation \cite{twc98:10a} of $ q(x,A;D) $ show that
$ \gamma $ is independent of $ r $ and equal to $ e^2/(32 \pi^2 ) $.
Then Eq. (\ref{eq:index_thm_a}) becomes
\bea
\frac{e^2}{32\pi^2} \ c[ D_c ] \ \sum_x
\epsilon_{\mu\nu\lambda\sigma} F_{\mu\nu} (x)
                               F_{\lambda\sigma}(x+\hat\mu+\hat\nu)
= N_{-}[D_c,A] - N_{+}[D_c,A]
\label{eq:index_theorem}
\eea
where
\beq
c[D_c]  = \left\{  \begin{array}{ll}
                   1   &  \mbox{ if $ D_c $ is topologically proper }       \\
                   0   &  \mbox{ if $ D_c $ is topologically trivial   }    \\
\mbox{integer} \ne 0, 1  &  \mbox{ if $ D_c $ is topologically improper   } \\
                    \end{array}
              \right.
\label{eq:cDc}
\eeq
{\it For any chirally symmetric $ D_c $ satisfying constraints (a)-(e),
Eq. (\ref{eq:index_theorem}) is the exact realization of the Atiyah-Singer
index theorem on the lattice and it holds for any Ginsparg-Wilson fermion
operator $ D = D_c ( \Id + R D_c )^{-1} $. }

Although the explicit form of the current $ G_\mu(x,A;D) $ in
Eq. (\ref{eq:qFkR}) is supposed to be very complicated, however, it is
instructive to compare Eq. (\ref{eq:qFkR}) to the divergence of axial
vector current on the lattice \cite{gwr,ph98:1},
\bea
 \partial_\mu \langle J^{5}_{\mu}(x,A;D) \rangle =  2 \ q(x,A;D)
  + 2 \sum_{s}^{N_{+}} [\phi^s_{+}(x)]^{\dagger} \phi^s_{+} (x) -
    2 \sum_{t}^{N_{-}} [\phi^t_{-}(x)]^{\dagger} \phi^t_{-} (x)
\label{eq:divJ5}
\eea
where $ \phi^s_{+} $ and $ \phi^t_{-} $ are normalized eigenfunctions
of $ D $ ( $ D_c $ ) with eigenvalues $ \lambda_s = \lambda_t = 0 $ and
chiralities $ +1 $ and $ -1 $ respectively.
Eliminating $ q(x,A;D) $ from both equations, we obtain
\bea
\partial_\mu G_\mu(x,A;D) &=&
\frac{1}{2} \partial_\mu \langle J^{5}_{\mu}(x,A;D) \rangle
 - \frac{e^2}{32\pi^2} \ c[ D ] \ \epsilon_{\mu\nu\lambda\sigma}
   F_{\mu\nu} (x) F_{\lambda\sigma}(x+\hat\mu+\hat\nu)       \nn
& & - \sum_{s}^{N_{+}} [\phi^s_{+}(x)]^{\dagger} \phi^s_{+} (x)
    + \sum_{t}^{N_{-}} [\phi^t_{-}(x)]^{\dagger} \phi^t_{-} (x)
\label{eq:Gmu}
\eea
It is interesting to note that all dependences on $ R $ are in the currents
$ G_\mu $ and $ J^5_\mu $. The zero modes and the $ F \tilde{F} $ term do not
depend on $ R $ as we have discussed above. Then we obtain
\bea
\delta_R \ \partial_\mu \
\left[ G_\mu(x,A;D) - \frac{1}{2} \langle J^{5}_{\mu}(x,A;D) \rangle \right]=0
\label{eq:GJ5}
\eea
This equation provides a $R$-invariant constraint between the current \\
$ G_\mu(x,A;D) $ and the axial vector current $ J^5_\mu(x,A;D) $.
The axial vector current $ J^{5}_{\mu}(x,A;D) $ is defined \cite{gwr}
in terms of the kernel $ K^5_{\mu}(x,y,z,A;D) $ through the equation
\bea
J^{5}_{\mu}(x,A;D) = \sum_{y,z} \bpsi_{x+y} K^5_{\mu}(x,y,z,A;D) \psi_{x+z}
\eea
where $ K^5_{\mu} $ is expressed in terms of the kernel
$ K_\mu $ for the vector current,
\bea
K^5_{\mu} = \frac{1}{2} ( K_{\mu} \gamma_5 - \gamma_5 K_{\mu} ).
\label{eq:K5}
\eea
Ginsparg and Wilson \cite{gwr} proved that the kernel for the vector current \\
$ K_\mu(x,y,z,A;D) $ is equal to $ D ( x+y, x+z, A ) $ times
the sign of $ (y-z)_\mu $ and times the fraction of the shortest
length paths from $x+z$ to $x+y$ which pass through the link from
$x-\hat\mu$ to $x$.
Finally the axial anomaly in Eq. (\ref{eq:qFkR}) becomes
\bea
q(x,A;D) = \frac{e^2}{32 \pi^2 } \ c[D_c] \
           \epsilon_{\mu\nu\lambda\sigma} F_{\mu\nu} (x)
                    F_{\lambda\sigma}(x+\hat\mu+\hat\nu)
           + \partial_\mu G_\mu(x,A;D)
\label{eq:q_final}
\eea
where $ c[D_c] $ is defined in Eq. (\ref{eq:cDc}) and the divergence term
$ \partial_\mu G_\mu(x,A;D) $ is related to the divergence of
axial vector current by Eqs. (\ref{eq:Gmu}) and (\ref{eq:GJ5}).

To summarize, we have clarified the role of the Ginsparg-Wilson relation
in the realization of the axial anomaly and the exact Atiyah-Singer index
theorem on the lattice. The crucial point is the existence of a
topologically proper chirally symmetric Dirac operator $ D_c $ in the
chiral limit where the Ginsparg-Wilson relation is turned off. If $ D_c $
satisfies the constraints (a)-(e) discussed in this paper, then any
Ginsparg-Wilson fermion operator $ D $ constructed by Eq. (\ref{eq:gwf})
could reproduce the correct continuum physics, in particular, the correct
axial anomaly and the exact index theorem. From this point of view, the
underlying reason why Neuberger-Dirac fermion really works is because the
existence of the topologically proper $ D_c $ ( or $ V $ )
[ Eq. (\ref{eq:DcV}) ] which stems from the Overlap formalism
\cite{rn95,hn98:8}. The Ginsparg-Wilson relation only provides
a scheme to break the continuum chiral symmetry in such a delicate way that
the exact chiral symmetry on the lattice is preserved and the axial anomaly
and the index are invariant while smooth deformations of $ R $ away from
zero bring the non-local and sometimes singular $ D_c $ into a class of
local and well defined Ginsparg-Wilson fermion operators $ D $.
The most crucial and highly non-trivial task is to find a `good seed'
$ D_c $ which at least satisfies all constraints (a)-(e).
So far there is only one good $ D_c $ found {\it explicitly}.
However, there is no indications that $ D_c $
is unique and thus the search for computationally less expensive
$ D_c $ is highly desirable. Since the Overlap formalism did
provide vital insights in discovering a good $ D_c $ \cite{hn97:7},
it would be interesting to see whether it could guide us to find
a better one. Mathematically, it is interesting to understand what
actually determines the topological characteristics of an operator
and how to construct a topologically proper operator in a systematic way.
For the axial anomaly (\ref{eq:qFkR}) recently constructed by
L\"uscher \cite{ml98:8}, we show that
the $ F \tilde{F} $ term is $ R $-invariant and the proportional
constant is $ e^2/(32 \pi^2 ) $ times $ c[D_c] $ which only
depends on the topological characteristics of $ D_c $; and the
divergence term is related to the divergence of axial vector current
by an $ R $-invariant constraint, Eqs. (\ref{eq:Gmu}) and (\ref{eq:GJ5}).
For any chirally symmetric $ D_c $ satisfying constraints (a)-(e),
the Atiyah-Singer index theorem can be realized exactly on the lattice,
Eq. (\ref{eq:index_theorem}), and it holds for any Ginsparg-Wilson fermion
operator $ D = D_c ( \Id + R D_c )^{-1} $.

\bigskip
\bigskip

\eject

\flushpar
{\bf Acknowledgement }
\bigskip

\noindent
I would like to thank Martin L\"uscher and Herbert Neuberger
for enlightening correspondences. I also wish to thank
Sergei V. Zenkin for many discussions in the last few months.
This work was supported by the National Science Council, R.O.C.
under the grant number NSC88-2112-M002-016.

\vfill\eject

\vfill\eject

\end{document}